\documentclass[aps,prl,twocolumn,superscriptaddress,showpacs]{revtex4}
\usepackage{amsfonts}
\usepackage{amssymb}
\usepackage{amsmath}
\usepackage{graphicx}

\usepackage{dcolumn}   
\usepackage{bm}        
\usepackage{flafter}

\begin{document}


\title{Superconductivity in Heavy Alkaline-Earths Intercalated Graphites}

\date{\today}

\author{J. S. Kim}
\affiliation{Max-Planck-Institut f\"{u}r Festk\"{o}rperforschung,
Heisenbergstra$\rm\beta$e 1, D-70569 Stuttgart, Germany}
\author{L. Boeri}
\affiliation{Max-Planck-Institut f\"{u}r Festk\"{o}rperforschung,
Heisenbergstra$\rm\beta$e 1, D-70569 Stuttgart, Germany}
\author{J. R. O'Brien}
\affiliation{Quantum Design, 6325 Lusk Boulevard, San Diego,
California 92121, USA}
\author{F. S. Razavi}
\affiliation{Department of Physics, Brock University, St.
Catharines, Ontario, L2S 3A1, Canada}
\author{R. K. Kremer}
\affiliation{Max-Planck-Institut f\"{u}r Festk\"{o}rperforschung,
Heisenbergstra$\rm\beta$e 1, D-70569 Stuttgart, Germany}

\begin{abstract}
We report the discovery of superconductivity below 1.65(6) K in
Sr-intercalated graphite SrC$_6$, by susceptibility and specific
heat ($C_p$) measurements. In comparison with CaC$_6$, we found
that the anisotropy of the upper critical fields for SrC$_6$ is
much reduced. The $C_p$ anomaly at $T_c$ is smaller than the BCS
prediction indicating an anisotropic superconducting gap for
SrC$_6$ similar to CaC$_6$. The significantly lower $T_c$ of
SrC$_6$ as compared to CaC$_6$ can be understood in terms of
"negative" pressure effects, which decreases the electron-phonon
coupling for $both$ in-plane intercalant and the out-of-plane C
phonon modes. We observed no superconductivity for BaC$_6$ down to
0.3~K.

\end{abstract}
\smallskip


\maketitle

The discovery of superconductivity in YbC$_6$ and CaC$_6$
\cite{YbC6:weller:syn,CaC6:emery:syn} initiated intensive
theoretical and experimental investigations
\cite{GIC:csanyi:band,GIC:mazin:band,CaC6:calandra:band,CaC6:lamura:penetration,CaC6:jskim:Cp,CaC6:smith:pressure,CaC6:jskim:pressure,CaC6:gauzzi:pressure,CaC6:bergeal:STM,CaC6:hinks:isotope}
on the alkaline-earth graphite intercalation compounds (GICs). The
results of the experimental studies, specially the observation of
a Ca isotope effect \cite{CaC6:hinks:isotope}, strongly favor
electron-phonon($e$-ph) coupling rather than exotic electronic
mechanisms \cite{GIC:csanyi:band}. Recent \emph{ab-initio}
electronic structure calculations prove that, in contrast to
MgB$_2$, $e$-ph coupling involving electronic interlayer (IL)
states becomes relevant \cite{GIC:csanyi:band} and is sufficiently
strong to generate the relatively high $T_c$'s
\cite{GIC:mazin:band,CaC6:calandra:band,CaC6:jskim:pressure}.
These findings stimulated discussions on the variability of the
\textit{e-ph} coupling strength in different branches of the
electronic and vibrational states for honeycomb layered compounds
like MgB$_2$, CaSi$_2$\cite{CaSi2:sanfilippo:Tc}, (Ca,Sr)AlSi
\cite{CaAlSi:mazin:band} and also the hypothetical Li$_2$B$_2$
\cite{Li2B2:kolmogorov:band}.

Many issues still remain open, especially about the nature of the
relevant phonons. In order to modify the phonon spectrum, Ca
isotope substitution \cite{CaC6:hinks:isotope} and hydrostatic
pressure experiments
\cite{CaC6:jskim:pressure,CaC6:gauzzi:pressure,CaC6:smith:pressure}
have been performed. The isotope experiments reported a
surprisingly high isotope exponent for Ca, $\alpha$(Ca) $\approx$
0.5, close to the BCS limit, suggesting a dominant role of the Ca
phonons in the $e$-ph coupling \cite{CaC6:hinks:isotope}. However,
\emph{ab-initio} calculations predicted similar isotope exponents
$\sim$ 0.25 for Ca and C, pointing to comparable contributions to
the $e$-ph coupling from the Ca and C phonons
\cite{CaC6:calandra:band}. The positive pressure dependence of
$T_c$ found for CaC$_6$ has been discussed in terms of phonon
softening for the in-plane Ca vibrations
\cite{CaC6:jskim:pressure,GIC:calandra:band}. Although the
experimentally determined $T_c$'s grow almost linearly with
pressure, the \textit{ab-initio} calculations predicted a
non-linear increase with a reduced magnitude
\cite{CaC6:jskim:pressure}. To resolve these discrepancies,
possible anharmonic effects of the ultrasoft intercalant phonon
modes, or a continuous superconducting gap distribution due to
anisotropic $e$-ph coupling have been suggested
\cite{GICs:Mazin:review,CaC6:sanna:band}, asking for further
investigations.

Another way to modify the relevant phonon modes is to vary the
intercalant species and replace Ca with other alkaline-earths such
as Sr or Ba. Mazin pointed out that for CaC$_6$ and YbC$_6$ the
square root of the mass ratio of the intercalants is only 15\%
larger than the ratio of their $T_c$'s \cite{GIC:mazin:band}.
Thus, according to this "isotope" effect argument other
alkaline-earths GICs may as well be superconducting. In fact,
subsequent \textit{ab-initio} calculations predicted
superconductivity for SrC$_6$ at 3.1 K and BaC$_6$ at 0.2
K\cite{GIC:calandra:band}. In this Letter, we report the discovery
of superconductivity in SrC$_6$ at $T_c$ = 1.65(6)~K by
susceptibility and specific heat measurements and the absence of
superconductivity in BaC$_6$ down to $\sim$ 0.3~K. The
superconducting properties of SrC$_6$ as well as the
\textit{ab-initio} calculations clearly demonstrate that SrC$_6$
can serve to bridge the two seemingly different classes of the
superconducting GICs: the low-$T_c$ alkali GICs and the
newly-discovered "high-$T_c$" systems, CaC$_6$ or YbC$_6$.
Furthermore, the comparison of SrC$_6$ with CaC$_6$ provides a
better insight into the unconventional nature of superconductivity
in alkaline-earth GICs.

Samples of SrC$_6$ and BaC$_6$ were synthesized from pieces of
highly oriented pyrolytic graphite (Advanced Ceramics, size
$\approx$ 3$\times$1$\times$1 mm$^3$) and Sr (99.95\%) or Ba
(99.95\%) metal by a vapor phase reaction performed for more than
a month at 470$^{\rm o}$C and 500$^{\rm o}$C, respectively.  X-ray
diffraction patterns show no reflections due to pristine graphite
or other higher stage intercalated phases, indicating good sample
homogeneity.  The graphite layer distance increases from CaC$_6$
(4.50~$\rm\AA$) to SrC$_6$ (4.95~$\rm\AA$) and to BaC$_6$
(5.25~$\rm\AA$) as expected from the size of the intercalant
atoms.  The stacking sequence of SrC$_6$ and BaC$_6$ is found to
be of the $\alpha\beta$-type (space group P6$_3$/mmc)
\cite{SrC6:guerard:syn} which differs from the
$\alpha\beta\gamma$-type stacking in CaC$_6$. Superconductivity
for SrC$_6$ was first observed by $ac$ magnetic susceptibility
(119 Hz) measurements to 0.3 K, and subsequently confirmed by
specific heat measurements using a PPMS $^3$He calorimeter
(Quantum Design). The pressure dependence of $T_c$ was measured up
to $\sim$ 1 GPa as described in detail
elsewhere\cite{CaC6:jskim:pressure}. In order to understand the
superconducting properties of SrC$_6$ and compare with those of
CaC$_6$, we also performed \textit{ab-initio} calculations of the
electron-phonon properties for SrC$_6$ using the experimental
$\alpha\beta$ stacking~\cite{PWscf,note:abinitio}.

Figure 1 shows a sharp superconducting transition in the magnetic
susceptibility $\chi(T)$ at $T_c$~=~1.65~(6) K in SrC$_6$. The
transition width $\Delta~T_c$$\approx$ 0.06 K was determined as
the temperature difference between 10$\%$ and 90$\%$ of the
diamagnetic shielding. In a magnetic field, $T_c$ shifts to lower
temperatures and the superconducting transition broadens. In
BaC$_6$, we cannot find any signature of superconductivity down to
0.3 K.

\begin{figure}
\includegraphics[width=6.0cm,bb=10 10 205 220]{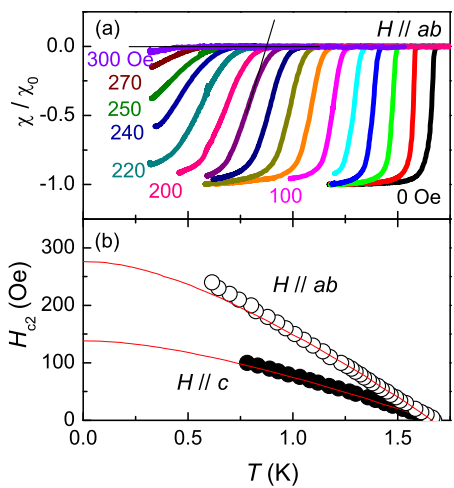}
\vspace{-3mm}
\caption{\label{fig1}(Color online) (a) Temperature dependence of
the $ac$ susceptibility of SrC$_6$ for various fields (as
indicated) perpendicular to the $c$-axis. $T_c$ is determined from
the intersection between the extrapolated lines of the steepest
slope of $\chi(T)$ and of the normal state $\chi(T)$ as shown by
the solid (black) lines. (b) The $H_{c2}$($T$) for $H$ $\parallel$
$ab$ plane and $H$ $\parallel$ $c$ axis. The WHH curve for both
$H$ directions are shown with (red) lines.}
\end{figure}

The upper critical fields, $H_{\rm c2}^{\rm{\parallel},\perp}$,
for $H$ parallel and perpedicular to the $c$ axis follow the
Werthamer-Helfand-Hohenberg (WHH) prediction  rather well
\cite{Hc2:WHH}(Fig.~\ref{fig1}(b)). A slight deviation from the
WHH curve is found at low temperatures, which has been observed in
other GICs such as CaC$_6$ and KC$_8$ \cite{KC8:koike:Hc2}. Within
the scope of the WHH approximation, we obtained
$H_{c2}^{\parallel}$(0) $\approx$ 138 Oe and $H_{c2}^{\perp}$(0)
$\approx$ 276 Oe as well as the corresponding coherence lengths,
$\xi_{ab}$(0) $\approx$ 1510 ${\rm \AA}$ and $\xi_{c}$(0)
$\approx$ 700 ${\rm \AA}$. Similar to CaC$_6$, $\xi_{c}$(0) is
larger than the graphite layer distance, indicating 3-dimensional
(3D) superconductivity.

\begin{figure}
\includegraphics[width=6.5cm,bb=10 10 230 210]{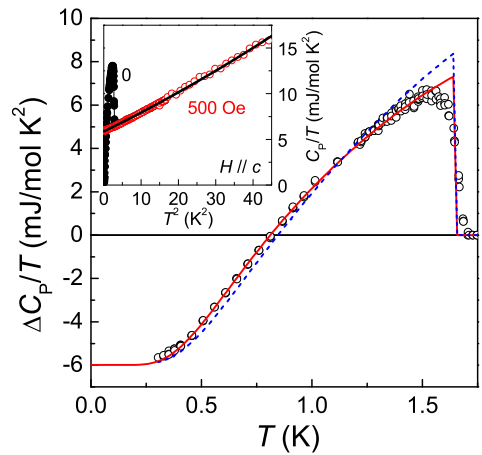}
\vspace{-3mm}
\caption{\label{fig2}(Color online) Temperature
dependence of $\Delta$$C_p$/$T$ = $C_p$($H$= 0)/$T$ - $C_p$($H$ =
500 Oe)/$T$. The (blue) dashed and (red) solid lines represent the
BCS curve and the best fit according to the $\alpha$-model (see
text), respectively. The inset shows the temperature dependence of
$C_p$ at $H$= 0 and 500 Oe. The solid (black) line through the
data points for $H$ = 500 Oe is a fit to a polynomial as described
in the text.}
\end{figure}

Bulk superconductivity in SrC$_6$ is confirmed by specific heat
($C_p$) measurements. The characteristic $C_p$ anomaly at $T_c$ =
1.65 K  is clearly observed at $H$ = 0, and completely disappears
with $H$ = 500 Oe ($H > H_{\rm c2}$(0)). There is no offset of
$C_p$/$T$ at $H$= 0 as $T$ $\rightarrow$ 0 K, indicating a
complete superconducting phase. Similar to CaC$_6$, the normal
state $C_p$ deviates slightly from a $T^3$ dependence (the inset
of Fig.~\ref{fig2}), due to low-lying Einstein phonon modes. The
normal state $C_p$ can well be described by $C_p$($T$) =
$\gamma_N$$T$ + $C_{ph}$($T$) where $\gamma_N$ is the Sommerfeld
constant, and $C_{ph}$($T$) = $\beta$$T^3$ + $\delta$$T^5$ is the
lattice contribution. The best fit to the $H$ = 500 Oe data yields
$\gamma_N$ = 5.92(1) mJ/mol K$^2$, $\beta$ = 0.191(1) mJ/mol
K$^4$, and $\delta$ = 0.801(4) $\mu$J/mol K$^6$. The estimated
Debye temperature $\Theta_D$(0) = 414(1) K is lower than that of
CaC$_6$ ($\Theta_D$(0) = 598 K) as expected from the atomic mass
difference. From a comparison of $\gamma_N$ with the calculated
density of states at the Fermi level $E_F$, $N(0)$, we estimate
the $e$-ph coupling strength $\lambda$ using the relation
$\gamma_N$ = (2$\pi^2$$k_B^2$/3)$N$(0)(1+$\lambda$). With $N(0)$ =
1.63 states/eV$\cdot$cell, we arrive at $\lambda$ = 0.54(1),
somewhat lower than in CaC$_6$, but still in the intermediate
coupling regime.

The difference $\Delta C_p$ between the normal and the
superconducting state is shown in Fig.~\ref{fig2}. At low
temperatures $\Delta C_p(T)$ exceeds the BCS prediction while it
is slightly lower than the BCS value near $T_c$. Using the
`$\alpha$-model' which assumes an isotropic $s$-wave BCS gap
$\Delta$($T$) scaled by the factor $\alpha$ =
$\Delta$(0)/$k_B$$T_c$, we were able to fit the detailed
temperature dependence of $\Delta C_p(T)$/$T$ by adjusting the gap
ratio to $\alpha$ = 1.67. Accordingly, the corresponding
2$\Delta$(0)/$k_B$$T_c$ = 3.34 is somewhat reduced below the weak
coupling BCS limit of 3.52.

First, we discuss the anisotropy of the superconducting properties
of SrC$_6$. The anisotropy of the upper critical fields,
$\it\Gamma_H$ = $H^{\perp}_{c2}$/$H^{\parallel}_{c2}$ amounts to
$\approx$ 2 at $T$ $\sim$ $T_c$/2, close to that of YbC$_6$
\cite{YbC6:weller:syn}, but significantly smaller than found in
CaC$_6$ ($\it\Gamma_H$ $\sim$ 4) \cite{CaC6:emery:syn}. Assuming
an isotropic superconducting gap, the anisotropy $\it\Gamma_H$
reflects the anisotropy of the Fermi velocities. Our
\textit{ab-initio} calculations for the Fermi surface (FS) of
CaC$_6$ and SrC$_6$ clearly reveal an elliptical sheet, associated
mainly to interlayer states, and a tubular structure, of $\pi^*$
character ($cf.$ Ref.~\cite{GIC:calandra:band}). In SrC$_6$ the FS
for the IL bands has a more pronounced 2D character and is open,
which results from the 10$\%$ increase of the $c$ axis lattice
parameter as compared to CaC$_6$. The  anisotropy of the average
Fermi velocities
$\it\Gamma_{v_F}$=$\left({<v^2_{F_{ab}}>}/{<v^2_{F_c}>}\right)^{1/2}$ is very
close in the two compounds: $\it\Gamma_{v_F}$ $\approx$ 1.9 in
CaC$_6$ and $\approx$ 1.7 in SrC$_6$. If we, however, consider
only the IL sheets where the superconducting gap is larger
\cite{CaC6:sanna:band}, we find a much larger difference:
$\it\Gamma_{v_F}$ = 1.1 for CaC$_6$ and $\it\Gamma_{v_F}$ = 2.1
for SrC$_6$. Therefore, $\it\Gamma_H$ of SrC$_6$ is expected to be
larger or at least similar to that of CaC$_6$. The significantly
enlarged anisotropy of $H_{c2}$ in CaC$_6$ with respect to SrC$_6$
therefore cannot be understood in terms of the anisotropy of the
Fermi velocities but must be attributed to an anisotropy of the
superconducting gap as well.

\begin{figure}
\includegraphics[width=6.5cm,bb=10 10 220 200]{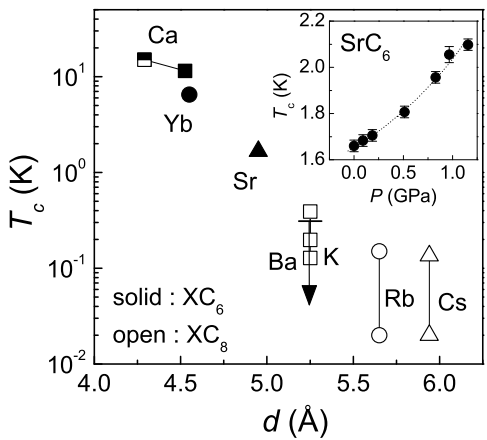}
\vspace{-3mm}
\caption{\label{fig3} $T_c$ as a function of the graphite layer
distance, $d$ for the alkali-GICs, $X$C$_8$ ($X$ = K, Rb and Cs)
and the alkaline earth-GICs $X$C$_6$ ($X$ = Ca, Yb, Sr, and Ba).
For CaC$_6$, $T_c$ at high pressure ($P$ = 8
GPa)\cite{CaC6:gauzzi:pressure} is also plotted (the half-shaded
square) and the graphite layer distance for the compressed CaC$_6$
is estimated from the theoretically calculated bulk modulus
\cite{CaC6:jskim:pressure}. The upper limit of $T_c$ for BaC$_6$
is indicated by the arrow. The inset shows $T_c$ vs. pressure for
SrC$_6$.}
\end{figure}

This conclusion is supported by the reduced $C_p$ jump observed in
SrC$_6$. Assuming an isotropic gap, the normalized jump,
$\Delta$$C_p$($T_c$)/$\gamma_N$$T_c$, grows with the $e$-ph
coupling strength over the BCS weak coupling limit of 1.432. With
$\lambda$ $\approx$ 0.54 estimated from $C_p$,
$\Delta$$C_P$($T_c$)/$\gamma_{N}$$T_c$ is expected to be enhanced
over the BCS value. The experimental
$\Delta$$C_p$($T_c$)/$\gamma_{N}$$T_c$ = 1.426, however, is
$smaller$ than the BCS value. As a characteristic feature of the
anisotropic superconducting gap, the entropy "lost" near $T_c$ is
transferred to lower temperatures
(Fig.~\ref{fig2})~\cite{CaC6:sanna:band}. These findings indicate
that the superconducting gap in CaC$_6$ as well as in SrC$_6$ has
a marked anisotropy, which is also supported by recent
calculations~\cite{CaC6:sanna:band}, showing that CaC$_6$
indeed exhibits a strongly \textbf{k}-dependent superconducting
gap due to anisotropic $e$-ph interaction. Considering that the
deviation of $\Delta C_p$($T$) from the predicted curve for the
isotropic gap model is less pronounced for SrC$_6$ than for
CaC$_6$ \cite{GICs:Mazin:review}, the superconducting gap
anisotropy is weaker for SrC$_6$. This is also consistent with the
reduced anisotropy in $H_{c2}$. Therefore, replacing Ca with Sr
decreases not only the strength of the $e$-ph coupling, but also
its anisotropy.

We can now discuss the reduced $T_c$ of SrC$_6$, which is about
an order of magnitude smaller than $T_c$ of CaC$_6$. When Ca is
replaced with Sr, $T_c$ decreases much more than it would be
expected in the view of the alkaline-earth "isotopic"
substitution~\cite{GIC:mazin:band}. With $\sqrt{M_{\rm Sr}/M_{\rm
Ca}}\simeq 1.48$, one would expect a $T_c$ of 7.8 K for SrC$_6$,
hence  much larger than the experimental value. We thus
conclude that the mass of the intercalant is not the main factor
which determines $T_c$. Instead, we found that the $T_c$'s of the
superconducting GICs strongly depend on the graphite interlayer
distance, $d$. Figure~\ref{fig3} illustrates that $T_c$ decreases
rapidly, almost exponentially with increasing $d$ for both alkali-
and alkaline earth-GICs~\cite{GIC:dresselhaus:review,note:Tcs}.
The increase of $T_c$ with pressure found for
CaC$_6$~\cite{CaC6:jskim:pressure,CaC6:smith:pressure,CaC6:gauzzi:pressure}
and SrC$_6$ ($dT_c/dP$ $\sim$ 0.35 K/GPa, the inset of Fig.
~\ref{fig3}) clearly manifests a similar trend. Obviously, the
main factor that governs the significant decrease of $T_c$ in
SrC$_6$ as well as the absence of superconductivity in BaC$_6$
down to 0.3 K, is the increased distance between the graphite
layers.

\begin{figure}
\vspace{-3mm}
\includegraphics*[width=7.5cm, bb=10 10 225 105]{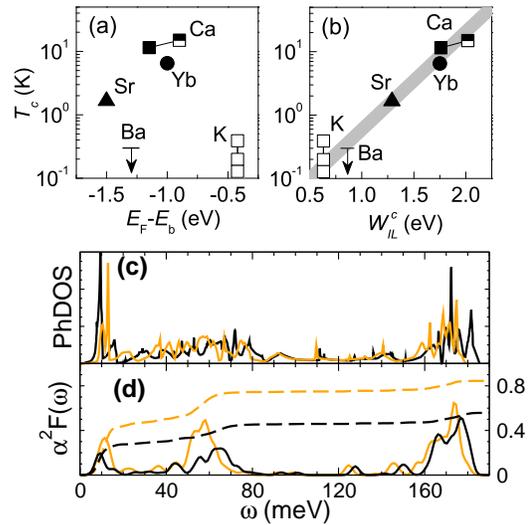}
\includegraphics*[width=6.5cm]{fig4cd.eps}
\vspace{-3mm}
\caption{\label{fig4}(Color online) (a) $T_c$ vs. the position of
the bottom of the interlayer bands ($E_b$) with respect to $E_F$
(b) $T_c$ vs. the band width of the interlayer bands along the $c$
axis ($W_{IL}^c$). The grey line is meant as guide to the eye. (c) Phonon
density of states and (d) Eliashberg function $\alpha^2 F(\Omega)$
and frequency-dependent electron-phonon coupling
$\lambda(\omega)=2\int_{0}^{\omega} \alpha^2 F(\Omega)/\Omega
d\Omega$ for CaC$_6$ (orange) and SrC$_6$ (black).}
\end{figure}

Figure~\ref{fig4} clearly demonstrates why. In the calculated
phonon density of states (PhDOS) (Fig.~\ref{fig4}(c)) and the
corresponding Eliashberg functions $\alpha^2 F(\omega)$
(Fig.~\ref{fig4}(d)) for CaC$_6$ and SrC$_6$, we observe three
groups of phonons: intercalant-related vibrations ($I_{xy}$ and
$I_z$) at $\omega$~$\le$~20~meV, C out-of-plane vibrations (C$_z$)
around 50~meV, and C bond-stretching vibrations (C$_{xy}$) at
$\omega$~$>$~150~meV. The qualitative shape of $\alpha^2
F(\omega)$ is the same in the two compounds, indicating a similar
spectral distribution of the $e$-ph interaction. But the total
$\lambda$ decreases from 0.83 in CaC$_6$ to 0.56 in SrC$_6$, while
the logarithmic-averaged phonon frequency ($\omega_{ln}$) remains
unchanged at $\sim$ 305 K. The results are in very good agreement
with the previous calculations for SrC$_6$ with the
$\alpha\beta\gamma$ stacking\cite{GIC:calandra:band}. Within
computational accuracy, the difference between the $\alpha\beta$
and $\alpha\beta\gamma$ stackings seems negligible in contrast to
a previous conjecture \cite{GIC:mazin:band}. Using the Allen-Dynes
formula, with $\mu^*$~=~0.145, we obtain $T_c$~=~11.4~K for
CaC$_6$ and $T_c$~=~3.1~K for SrC$_6$, in reasonable agreement
with the experimental results and previous
calculations\cite{GIC:calandra:band}.

The reduction of $T_c$ in SrC$_6$ is due to the simultaneous
decrease of the $I_{xy}$ and C$_{z}$ contribution to the $e$-ph
coupling. The reduced coupling for the low-lying $I_{xy}$
vibrations has a negative effect on $T_c$, but it is also very
effective in increasing $\omega_{ln}$, thus, the effect is partly
compensated. On the other hand, the large reduction of coupling
associated to C$_z$ vibrations, which happens at energy scales
comparable to $\omega_{ln}$, reduces $\lambda$ leaving
$\omega_{ln}$ unchanged, and thus has a large effect on the $T_c$.
The reduction of $e$-ph
coupling for C$_z$ vibrations in the intermediate energy range is
essential to explain the significantly lower $T_c$ for
SrC$_6$ than CaC$_6$.

The $e$-ph interaction for the  C$_{z}$ modes, and to a lesser
extent for the  $I_{xy}$ ones, is
associated to the interband coupling between the IL and $\pi^*$
states~\cite{GIC:boeri:band}.
Going from CaC$_6$ to SrC$_6$,
this interband coupling is essentially reduced
 by the increase of the
$c$ axis lattice constant, which decreases the real-space overlap
between IL and $\pi^*$ wave-functions. The characteristic
parameter monitoring the IL-$\pi^*$ overlap is the bandwidth of
the IL band along the $c$ axis ($W_{IL}^c$). Figure~\ref{fig4}(b)
demonstrates that $T_c$ for both alkali- and alkaline earth-GICs
is directly correlated to $W_{IL}^c$, clearly explaining the $T_c$
dependence on the graphite layer distance. This observation is in
strong contrast to the dependence of $T_c$ on the degree of
filling for the IL band (Fig.~\ref{fig4}(a)). Even though
this band has to be occupied for superconductivity to
occur~\cite{GIC:csanyi:band}, there is no clear correlation of
$T_c$ with the number of IL electrons.

Our findings indicate that C phonon modes play a nontrivial role
in the superconductivity of the GICs,
in contrast with the conjecture  from the Ca isotope
experiments~\cite{CaC6:hinks:isotope}.
Since C would give a sizable contribution to the isotope effect, the
total isotope exponent would exceed the BCS limit $\alpha_{T_c}=0.5$,
which requires further studies on the
effects of C isotope substitution.

In conclusion, we reported that SrC$_6$ becomes superconducting at
$T_c$ = 1.65(6) K, but BaC$_6$ stays normal conducting down to 0.3
K. The reduced $C_p$ jump at $T_c$ in SrC$_6$ and the reduced
anisotropy $H_{c2}$ in comparison with the corresponding data for
CaC$_6$ strongly supports the idea of an anisotropic
superconducting gap \cite{GICs:Mazin:review,CaC6:sanna:band}. We
give clear evidence that $T_c$ of the GICs essentially depends on
the graphite layer distance due to sensitive change of $e$-ph
coupling for $both$ in-plane intercalant and the out-of-plane C
phonon modes. Our results suggest that a possible route to
increase $T_c$ is to replace Ca by smaller atoms, such as Mg.
Several attempts to prepare pure MgC$_6$ have failed so far, but
partial substitution of Mg or Li, as long as the filling of the IL
bands is kept, could be possible to reduce $d$, stabilize the
structure, and as a result, increase $T_c$.

\acknowledgments The authors acknowledge useful discussion with A.
Simon, O. K. Andersen, G. B. Bachelet, M. Giantomassi and D.
Guerald, and we thank E. Br\"{u}cher, S. H\"{o}hn for experimental
assistance.


\begin{thebibliography}{}
\bibitem{note:abinitio}
For computational details, see Ref.~\cite{CaC6:jskim:pressure}.
For Sr, we used an ultrasoft \cite{Vanderbilt} pseudopotential,
with semicore $s$ and $p$ states treated as valence.
\bibitem{note:Tcs} Here we do not consider the metal-rich GICs synthesized under high pressures such as LiC$_2$ which are known to be $metastable$.
\end{thebibliography}
\end{document}